\documentclass[twocolumn,preprintnumbers,amsmath,amssymb]{revtex4}

\usepackage{graphicx}

\usepackage{dcolumn}

\usepackage{bm}

\begin{document}

\title{HIGHER DIMENSIONAL  METRICS  OF COLLIDING GRAVITATIONAL PLANE WAVES }

\author{ M. G{\" u}rses} 

\altaffiliation
{email:
    gurses@fen.bilkent.edu.tr}

\affiliation{Department of Mathematics, Faculty of Sciences,
 Bilkent University, 06533 Ankara - Turkey}

\author{E.O. Kahya }

\altaffiliation{email: kahya@metu.edu.tr}

\author{A. Karasu }

\altaffiliation
{email: karasu@metu.edu.tr}

\affiliation{
 Department of Physics, Faculty of Arts and  Sciences,
Middle East Technical University, 06531 Ankara-Turkey}%


\begin{abstract}

We give  a higher even dimensional extension of vacuum 
colliding gravitational plane waves  with the combinations of collinear
and non-collinear polarized four-dimensional metric. The
singularity  structure of space-time  depends on the parameters of the solution.

\end{abstract}


\maketitle

\section{INTRODUCTION}

One of the main fields of  interest in general relativity is the collision
of gravitational plane waves. The structure of the field equations, physical and
 geometrical interpretations, and various solution-generating
 techniques have been described in \cite{gri}.
 Khan-Penrose \cite{kp} and Szekeres \cite{szk}
have found exact solutions of the vacuum Einstein equations describing 
 the collision of impulsive and shock plane waves with  collinear polarizations.
Nutku-Halil \cite{nh} generalized the Khan-Penrose metric
to the case of non-collinear  polarizations. Later several authors 
have studied exact solutions of the  Einstein-vacuum and   Einstein-Maxwell equations,
describing the collision of the gravitational and electromagnetic plane
waves. In general relativity, various techniques are known for generating  different solutions of vacuum 
and electrovacuum
Einstein field equations  \cite{krm}. In this context
recently various new
solution-generating techniques have been given  for vacuum and electrovacuum cases
 \cite{alx,alx1}.  In the low energy limit of string theory the colliding
gravitational plane waves and some exact solutions  are given in \cite{gur,gar}.  Also a formulation of
the colliding gravitational  plane waves in metric-affine theories  is given in \cite{gar1}.
In connection
with  string theory colliding gravitational plane waves were studied in 
\cite{das}- \cite{Bo}.

 Recently,  motivated by the results obtained  in \cite{gur1,gur2},
 we showed that starting from a Ricci flat
metric of a four-dimensional geometry admitting two Killing vector fields it is possible to generate a whole
class $2N=2+2n$-dimensional  Ricci flat metrics \cite{gur3}.
As an explicit example we constructed  higher even
dimensional metrics of colliding gravitational waves from the corresponding four dimensional vacuum  Szekeres metrics.

 In this paper  we give a full construction of  higher even dimensional colliding 
gravitational plane waves  with the combinations of 
collinear and non-collinear polarized four-dimensional metrics .
 In particular,  we  show that there is no  higher even dimensional
solution for the Nutku-Halil solution. The singularity structure of this higher dimensional solutions is also
 examined by using the curvature invariant.

\section{HIGHER DIMENSIONAL COLLIDING GRAVITATIONAL PLANE WAVE GEOMETRIES}

In \cite{gur3} we have studied some Ricci flat geometries with an arbitrary signature.
We presented a procedure  to construct solutions to some  higher even dimensional Ricci flat
metrics.
According to our theorem stated in \cite{gur3}, 
 if  the metric functions ${\cal{U}}(x^{a}), \, h_{o\, i}(x^{a})$,  and ${\cal{M}}_{i}(x^{a})$ , for
each $i=1,2, \cdots, n$, form a  solution to the four dimensional  vacuum field equations for the metric

\begin{equation}
ds^2=e^{-{\cal{M}}_{i}}\, \eta_{a b}\,dx^{a}\,dx^{b}+
 e^{{\cal{U}}}\,(h_{0\,i})_{\,ab}\,dy^a~dy^b, 
\end{equation}

\noindent
where $\eta_{ab}$ is the metric of the flat 2-geometry with an arbitrary signature ($ 0$  or $\pm 2$),
then the metric of the $2N=2+2n$ dimensional geometry defined below 

\begin{equation}
ds^2=e^{-M}\,\eta_{a b}\, dx^{a}\,dx^{b}+ 
\sum_{i=1}^{n}\, \epsilon_{i}\,e^{u_{i}}\,(h_{0\,i})_{ab}\,dy_{i}^a\,dy_{i}^b,
\end{equation}

\noindent
solves the vacuum  equations, where 
$\epsilon_{i}=\pm 1$, $M=\bar M+\tilde M $, $\tilde M=
\sum_{i=1}^{n}\, {\cal{M}}_{i}$ . $\bar M$ solves

\begin{eqnarray}
\frac{1}{2} (\bigtriangledown_{\eta}^{2} \bar M)\, \eta_{ab}
+(n-1)\,{\cal{U}}_{,ab}-
\frac{1}{2} [\bar M_{,a} \, {\cal{U}}_{,b} 
+\bar M_{,b}\, {\cal{U}}_{,a} \nonumber\\
-\bar M_{,d} \, {\cal{U}}_{,}^{d}\, \eta_{ab}]
-\frac{1}{2} \sum_{i=1}^{n}\partial_{a}\,u_{i}\partial_{b}\,u_{i}
+ \frac{1}{2} n \partial_{a}\, {\cal{U}}\partial_{b}\, {\cal{U}}
=0, \label{bb}
\end{eqnarray}
and  $\cal {U}$  and $u_{i}$  solve the following equations,  respectively:
\begin{eqnarray}
&&\partial_{a}[\eta^{ab}\,e^{{\cal{U}}} \partial_{b} {\cal{U}}]=0, \label{bb1}\\
&&\nabla^{2}_{\eta}\, u_{i}+\eta^{ab}\, {\cal{U}}_{,a}\,u_{i,b}=0. \label{bb2}
\end{eqnarray}

\noindent
Here
the local coordinates of the ($2n+2$) dimensional geometry are given by
$x^{\alpha}=(x^{a}, y_{1}^{a}, y_{2}^{a}, \cdots, y_{n}^{a})$.
Given any four-dimensional metric of colliding vacuum gravitational
plane wave geometry we have their extensions to  higher dimensions for
arbitrary $2N$ without solving further differential equations.  In particular,
taking $u_{i}=m_{i} \cal{U}$, where $m_{i}  (i=1,2,...,n) $ are real constants
satisfying 
\begin{equation}
\sum_{i=1}^{n}\, m_{i}=1, ~~~~~~~\sum_{i=1}^{n}\, (m_{i})^{2}=m^{2},
\end{equation}

\noindent
and the signature of flat-space metric with null coordinates  is 
$$\eta=\left(\begin{array}{cc} 0&1\\1&0 \end{array}\;
\right),~~~ x^{1}=u,\, x^{2}=v,$$
then the  solutions to the  Eqs.~(\ref{bb}) and (\ref{bb1}) are found to be 

\begin{eqnarray}
e^{-M}&=&(f_{u}\,g_{v})^{-n+1}\,(f+g)^{\frac{m^2+n-2}{2}}
e^{-\sum_{i=1}^{n}\, {\cal {M}}_{i}}\label{mf}, \\
e^{\cal U}&=&f(u)+g(v),\label{ze}
\end{eqnarray}
where $f$ and $g$ are arbitrary functions of their arguments
and Eq.~(\ref{bb1}) is automatically satisfied as a result of Eq.~(\ref{ze}). 
Therefore, the above exact solutions describe the collision of gravitational
waves for arbitrary $n>1$.

In \cite{gur3} we  found a family of solutions when the four dimensional
metrics are the Szekeres metrics \cite{szk} (collinear 
four dimensional metrics). In the next section we generalize this solution by 
adding non-collinear metrics to the Szekeres metrics. We also show that,
in our formalism, there is no higher dimensional metric constructed
by the non-collinear four dimensional metric alone.

\section{HIGHER DIMENSIONAL VACUUM SOLUTIONS}

We take the following metric as the metric describing a plane wave  geometry  in $2N$ dimensions.
\begin{eqnarray}
ds^2&=&2e^{-M}\,du dv\nonumber \\+
&&\sum_{j=1}^{n_{a}}\, \,(f+g)^{m_{j}}\,\frac{|(1-E_{j})
  \,dx_{j}+i\,(1+E_{j})\,dy_{j})|^2}
{ 1-E_{j} \,{\bar E_{j}}} \nonumber \\
&+&\sum_{j=n_{a}+1}^{n}\,(f+g)^{m_{j}}\,(e^{V_{j}}\,dx_{j}^2+
e^{-V_{j}}\,dy_{j}^{2}),   \label{nh}
\end{eqnarray}
where  the  complex functions  $E_{i}$ (non-collinear case) and the real functions
$V_{i}$ (collinear case) satisfy
the Ernst and Euler-Poisson-Darboux  equations, respectively,
\begin{eqnarray}
&&(1-E_{i} {\bar E}_{i})\, \bigl [2(f+g)\,E_{i,\,fg}+E_{i,\,f}+E_{i,\,g} \bigr ] ~~~~\nonumber \\
&=&-4 (f+g){\bar  E}_{i}\,E_{i,\,f}\,E_{i,\,g}, ~~~~~~ i=1,2,...,n_{a},\label{er}
\end{eqnarray}
\begin{equation}
2(f+g) V_{i,\,fg}=-V_{i,\,f}- V_{i,\,g},~~~~~~~i=1,2,...,n-n_{a},\label{ep}
\end{equation}
with the following solutions:
\begin{equation}
E_{i}=e^{i\alpha_{i}} (\frac{1}{2}-f)^{1/2} (\frac{1}{2}+g)^{1/2}+e^{i\beta_{i} }(\frac{1}{2}+f)^{1/2} (\frac{1}{2}-g)^{1/2},
\end{equation}
\begin{equation}
V_{i}=-2k_{i}\, \tanh^{-1}\, \bigl (\frac{\frac{1}{2}-f}{\frac{1}{2}
+g} \bigr )^{\frac{1}{2}}-2 \ell_{i}\, \tanh^{-1}\, \bigl (\frac{\frac{1}{2}
 -g}{\frac{1}{2}
  +f} \bigr )^{\frac{1}{2}}, \label{vi}
\end{equation} 
where $ \alpha_{i}$, $\beta_{i}$, $k_{i}$, and $\ell_{i}$ are arbitrary constants and there is no sum on $i$ in
Eq.~(\ref{er}).
The metric function $M$ given in (\ref{nh}) is
\begin{eqnarray}
e^{-M}&=&(f_{u}\,g_{v})^{-n+1}\,(f+g)^{\frac{m^2+n-2}{2}}\nonumber\\
&& \times e^{-\sum_{i=1}^{n_{a}}\,{\cal{M}}^{(\text{1})}_{i}}
e^{-\sum_{i=n_{a}+1}^{n}\,{\cal M}^{(\text{2})}_{i}},
\end{eqnarray}
where  the metric functions  ${\cal {M}}^{(\text{1})}_{i}$    and   ${\cal {M}}^{(\text{2})}_{i}$  
are, respectively,
\begin{widetext}
\begin{eqnarray}
&&e^{-{\cal {M}}^{(\text{1})}_{i}}=\frac{f_{u}g_{v}[-\gamma_{i}^{2}(f+g)^{2}+2(\gamma_{i}^{2}-4)(1+4fg)fg+\frac{\gamma_{i}^{2}}{4}-1]}
{(f+g)[(1+4fg)+2\gamma_{i} (\frac{1}{4}-f^{2})^{1/2}(\frac{1}{4}-g^{2})^{1/2}]
 (\frac{1}{4}-f^{2})^{1/2}(\frac{1}{4}-g^{2})^{1/2}d_{i}}, \\
&&e^{-{\cal {M}}^{(\text{2})}_{i}}=\frac{c_{i}f_{u}g_{v}(f+g)^{\frac{\tau_{i}}{2}}}{(\frac{1}{2}-f)^{k_{i}^{2}/2}
(\frac{1}{2}-g)^{\ell_{i}^{2}/2}(\frac{1}{2}+f)^{\ell_{i}^{2}/2}(\frac{1}{2}+g)^{k_{i}^{2}/2}\,\,
 [(\frac{1}{2}-f)^{1/2}(\frac{1}{2}-g)^{1/2}+(\frac{1}{2}+f)^{1/2}(\frac{1}{2}+g)^{1/2}]^{2k_{i}\ell_{i}}},
\end{eqnarray}
\end{widetext}
where $\gamma_{i}=2\,\cos\,(\alpha_{i}-\beta_{i})$, $\tau_{i}={k_{i}^{2}+\ell_{i}^{2}+2k_{i}\ell_{i}-1}$,
 and   $d_{i}$, $ c_{i} $ are constants.

The metric function $e^{-M}$ must be continuous across the null boundaries.  To make it so we define
\begin{equation}
k^2 \equiv \sum_{i=1}^{n-n_{a}}\,k_{i}^{2},
~~~ \ell^2 \equiv \sum_{i=1}^{n-n_{a}}\,\ell_{i}^2, ~~~~ s \equiv \sum_{i=1}^{n-n_{a}}\, k_{i} \ell_{i},
\end{equation}
and we assume that the functions  $f$ and $g$ take the form
\begin{equation}
f=\frac{1}{2}-(e_{1}\,u)^{n_{1}} ,~~~g=\frac{1}{2}
  -(e_{2}v)^{n_{2}},
\end{equation}
where $e_{1},\,e_{2}$, $n_{1} \ge 2,~n_{2} \ge 2$ are real constants.
Then the metric function $e^{-M}$ is continuous across the boundaries if
\begin{equation}
k^{2}+n_{a}=2(1-\frac{1}{n_{1}}), ~~~~~~~~\ell^{2}+n_{a}=2(1-\frac{1}{n_{2}})\label{cn}
\end{equation}
with
\begin{equation}
1 \le k^2 <2,~~~ 1 \le \ell^2 < 2. \label{n1}
\end{equation}

Therefore, the metric function $e^{-M}$ reads
\begin{widetext}
\begin{equation}
e^{-M}=\frac{(f+g)^{\frac{\sigma
}{2}}}{(\frac{1}{2}+f)^{(n_{a}+\ell^{2})/2}(\frac{1}{2}+g)^{(n_{a}+k^{2})/2}}
[ (\frac{1}{(\frac{1}{2}-f)^{1/2}(\frac{1}{2}-g)^{1/2}+(\frac{1}{2}+f)^{1/2}(\frac{1}{2}+g)^{1/2}})^{2s}]
 \,\,e^{-\Gamma},
\end{equation}
\end{widetext}
\noindent
where
\begin{equation}
e^{-\Gamma}= \Pi_{i=1}^{n_{a}} \frac{\gamma_{i}^{2}(f+g)^{2}-2(\gamma_{i}^{2}-4)(1+4fg)fg-\frac{\gamma_{i}^{2}}{4}+1}{
(1+4fg)+2\gamma_{i}(\frac{1}{4}-f^{2})^{1/2}(\frac{1}{4}-g^{2})^{1/2}} \nonumber
\end{equation}
and  $\sigma=m^{2}+n-n_{a}+k^{2}+\ell^{2}+2s-3$.
We may set 
\begin{equation}
-2e_{1}e_{2}n_{1}n_{2} \Pi_{i=1}^{n-n_{a}} c_{i} =\Pi_{i=1}^{n_{a}} d_{i}.
\end{equation}
When we consider  only the non-collinear case  where $n_{a}=n$ and $k^{2}=\ell^{2}=s=0$,
 conditions in   Eq.~(\ref{cn}) imply 
$n_{1}=n_{2}$. Then  Eqs.~(\ref{cn}) and  (\ref{n1})  reduce to
\begin{equation} 
n=2(1-\frac{1}{n_{1}}),
\end{equation} 
where $n_{1} \ge 2$ and $n$ is a positive integer.
The only possible solution is $n=1$, $n_{1}=2$ which
corresponds to the Nutku-Halil solution. Hence if we take  all $E_{i}$ as the
 Nutku-Halil metric functions it is not possible to find
an appropriate metric function $M$ which is continuous across the null boundaries for $n>1$.
This is the reason why  one has to take the higher dimensional metric as the combinations 
of collinear and non-collinear polarizations.

\subsection{Singularity structure}

We now discuss the nature of the space-time
singularity. That is, we  study the behavior of the metric function $M$ as $f+g$ tends to zero.
For this purpose, using the result obtained in \cite{gur3} for the curvature invariant 
\begin{equation}
I=R^{\mu\nu\alpha\beta}R_{\mu\nu\alpha\beta},
\end{equation}
\noindent
as
\begin{equation}
I \sim e^{2M} \frac{(f_{u}g_{v})^{2}}{(f+g)^{4}},
\end{equation}
 we find
\begin{equation}
I \sim (f_{u}\,g_{v})^2\, (f+g)^{-\mu},
\end{equation}
as $f+g \rightarrow 0$. Here
 $\mu=k^2+\ell^2+m^2+2s+4n_{a}+2$.  For the four dimensional case ($n=1$)  
 $n_{a}=0$ with $k=\ell=s=1$ and $m^{2}=1$  this  corresponds to the singularity structure of the Khan-Penrose solution.
 $n_{a}=1$, $k^{2}=\ell^{2}=s=0$ and
$m^{2}=1$ corresponds to the Nutku-Halil solution.
 It is known that both solutions have the same singularity structures.
$n_{a}=0$, with $k=k_{1}$, $ \ell=\ell_{1}$, and $m^{2}=1$ corresponds to the singularity structure of the Szekeres solution. 
The singularity structure in the higher dimensional
spacetimes can be made weaker or stronger than the four dimensional cases by choosing the
constants $m_{i}, k_{i}$, and $\ell_{i}$ properly.

\section{conclusion}

In this work we gave a  higher even dimensional generalization  of vacuum colliding gravitational plane waves
with  the combinations of  collinear and non-collinear polarizations.
 We also discussed  the singularity structure
of the corresponding spacetimes, and showed that the strength of the singularity depends on
 arbitrary parameters $m_{i}$. We also showed  that it is not possible to construct the higher dimensional metric
by non-collinear four dimensional metric functions
 $E_{i}$ (\ref{er}) alone. Einstein's equations and continuity conditions force us to superpose
 the collinear and non-collinear
metric functions.

\vspace{1cm}

\begin{acknowledgments}

This work is partially supported by the Scientific and Technical
Research Council of Turkey (TUBITAK) and by Turkish Academy of Sciences (TUBA).

\end{acknowledgments}

\end{document}